\documentclass{elsart}

\usepackage{graphicx}

\usepackage{amssymb,amsmath,bm}
\usepackage{cite}
\usepackage{url}

\newtheorem{lemma}{Lemma}
\newtheorem{theorem}{Theorem}
\newtheorem{corollary}{Corollary}
\newtheorem{remark}{Remark}
\newenvironment{proof}{\noindent\textit{Proof}: }{\hfill$\blacksquare$\vskip 0.5\baselineskip}

\begin{document}

\begin{frontmatter}

\title{Cryptanalysis of a chaotic block cipher with external key and its improved version}
\thanks{A preliminary version of this paper has been presented at the 4th Regional
Inter-University Postgraduate Electrical and Electronics
Engineering Conference (Macau, China) in July, 2006.}

\author[hk-cityu]{Chengqing Li\corauthref{corr}},
\author[hk-polyu]{Shujun Li\corauthref{corr}},
\author[spain]{Gonzalo \'{A}lvarez},
\author[hk-cityu]{Guanrong Chen} and
\author[hk-polyu]{Kwok-Tung Lo}

\corauth[corr]{Corresponding authors: Chengqing Li
(cqli@ee.cityu.edu.hk), Shujun Li (http://www.hooklee.com).}

\address[hk-cityu]{Department of Electronic Engineering, City University of Hong Kong,
83 Tat Chee Avenue, Kowloon Tong, Hong Kong SAR, China}
\address[hk-polyu]{Department of Electronic and Information Engineering,
The Hong Kong Polytechnic University, Hung Hom, Kowloon, Hong Kong
SAR, China}
\address[spain]{Instituto de F\'{\i}sica Aplicada, Consejo Superior de
Investigaciones Cient\'{\i}ficas, Serrano 144, 28006 Madrid,
Spain}

\begin{abstract}
Recently, Pareek et al. proposed a symmetric key block cipher
using multiple one-dimensional chaotic maps. This paper reports
some new findings on the security problems of this kind of chaotic
cipher: 1) a number of weak keys exists; 2) some important
intermediate data of the cipher are not sufficiently random; 3)
the whole secret key can be broken by a known-plaintext attack
with only 120 consecutive known plain-bytes in one known
plaintext. In addition, it is pointed out that an improved version
of the chaotic cipher proposed by Wei et al. still suffers from
all the same security defects.
\end{abstract}
\end{frontmatter}

\section{Introduction}

Due to some close and subtle relation between statistical properties
of chaotic systems and cryptosystems, the idea of utilizing chaos to
design digital ciphers and analog secure communication schemes has
been attracting more and more attention during the past two decades
\cite{ShujunLi:Dissertation2003,
Yang:Survey:IJCC2004,AlvarezLi:Rules:IJBC2006}.

Since 2003, Pareek et al. proposed three different cryptosystems
based on one or more one-dimensional chaotic maps
\cite{Pareek:PLA2003, Pareek:CNSNS2005, Pareek:IVC2006}. Unlike most
existing chaotic ciphers \cite{ShujunLi:Dissertation2003}, in the
ciphers of Pareek et al., the initial conditions and/or the control
parameter are not used as the secret keys, but derived from an
external key instead, with the goal of obtaining a new way to
achieve a higher level of security. The chaotic ciphers proposed in
\cite{Pareek:PLA2003} and \cite{Pareek:CNSNS2005} have been
cryptanalyzed by Alvarez et al. in \cite{Alvarez:PLA2003}, and by
Wei et al. in \cite{JunWei:CNSNS2006}, respectively. Wei et al.
further proposed a remedy to improve the security of the original
cipher against known-plaintext attacks.

This paper re-examines the security of the chaotic cipher designed
in \cite{Pareek:CNSNS2005} and its improved version suggested in
\cite{JunWei:CNSNS2006}. Three new security problems of the original
cipher that were not reported in \cite{JunWei:CNSNS2006} are found:
1) there are a number of weak keys that cannot encrypt the
plaintexts at all; 2) some important intermediate data of the cipher
are not sufficiently random; 3) the secret key can be completely
broken by a known plaintext attack with only 120 consecutive known
plain-bytes in just one known-plaintext. In addition, it is found
that the improved cipher developed in \cite{JunWei:CNSNS2006} still
suffers from the same problems, thus failing to enhance the original
cipher's security.

The rest of the paper is organized as follows. The next section
gives a brief introduction to the original cipher of Pareek et al.
and its improved version. Section~\ref{sec:cryptanalysis} focuses
on the above-mentioned security problems of the two chaotic
ciphers under study. The last section concludes the paper.

\section{The Cipher of Pareek et al. and its Improved Version}
\label{sec:scheme}

In the original cipher of Pareek et al. \cite{Pareek:CNSNS2005},
the plaintext and the ciphertext are both arranged with 8-bit
blocks, i.e., arranged byte by byte as follows: $P=P_1P_2\cdots
P_n$ and $C=C_1C_2\cdots C_n$, where $P_i$, $C_i$ are the $i$-th
plain-byte and the $i$-th cipher-byte, respectively.

The secret key used in the cipher is a 128-bit integer $K$,
represented as $K=K_1K_2\cdots K_{16}$, where
$K_i\in\{0,1,\cdots,255\}$ which is called the $i$-th sub-key in
this paper\footnote{In \cite{Pareek:CNSNS2005}, $K_i$ is called
``session key". However, such a term may cause some confusion, since
``session keys" are generally used to denote randomly-generated keys
in cryptographical protocols.}. The secret key is used to generate
the initial conditions of four chaotic maps and the contents of two
dynamic tables. Then, each plain-byte is masked by the output of one
randomly-selected chaotic map after a number of iterations, under
the control of the two dynamic tables. After a group of plain-bytes
is encrypted, the two dynamic tables are updated following the
current chaotic state of the selected chaotic map. The number of
chaotic iterations and the group size are varying instead of being
fixed. More precisely, the chaotic cipher works as follows.

\begin{enumerate}
\item
The following four chaotic maps are marked with map number
$N=0,1,2,3$, respectively.

\begin{itemize}
\item
$N=0$ -- logistic map: $f(x)=\lambda x(1-x)$;

\item
$N=1$ -- tent map: $f(x)=\begin{cases}
\lambda x, & \mbox{if }x<0.5;\\
\lambda(1-x), & \mbox{if }x\leq 0.5;
\end{cases}$

\item
$N=2$ -- sine map: $f(x)=\lambda \sin(\pi x)$;

\item
$N=3$ -- cubic map: $f(x)=\lambda x(1-x^2)$.
\end{itemize}

In \cite{Pareek:CNSNS2005}, the control parameters of the above
four chaotic maps are assigned as $\lambda=3.99$, $\lambda=1.97$,
$\lambda=0.99$ and $\lambda=2.59$, respectively.

\item
The first dynamic table (DT1) stores the initial conditions (IC) of
the four chaotic maps. Before the encryption process starts, the
four initial conditions are all set to be the following
value\footnote{Note that we use an equivalent formula to replace
Eqs. (4) and (5) in \cite{Pareek:CNSNS2005}, trying to give a
clearer representation. Here ``mod 1'' means subtracting the integer
part and keeping only the fractional part, which lies in the
half-open interval $[0,1)$.}:

\begin{equation}\label{eq:generate_IC}
\mathrm{IC}=\left(\sum_{i=1}^{16}\frac{K_i}{256}\right)\bmod
1=\frac{\left(\sum_{i=1}^{16}K_i\right)\bmod 256}{256}.
\end{equation}

\item
Each entry of the second dynamic table (DT2) stores three distinct
values: the selected chaotic map that encrypts a group of
plain-bytes, the number of plain-bytes in a group that is encrypted
by the corresponding chaotic map, and the number of iterations of
the corresponding chaotic map for encrypting each plain-byte, which
are denoted by $N$, $B$ and IT, respectively. Given a linear
congruential pseudorandom number generator (LCG),

\begin{eqnarray}
Y_0 & = & \lfloor 100\times\mathrm{IC}\rfloor,\label{eq:LCG0}\\
Y_n & = & (5Y_{n-1}+1)\bmod 16,\mbox{ when }n\geq 1,\label{eq:LCG}
\end{eqnarray}
the three values of the $n$-th entry in DT2 are determined as
follows\footnote{Note that $Y_0$ is not confined in
$\{0,\cdots,15\}$, so it is just used as the seed of the LCG and
should not be considered as part of the LCG sequence to generate
the entries of DT2.}:
\begin{eqnarray}
N_n & = & Y_n\bmod 4,\label{eq:generate_N}\\
B_n & = & Y_n,\\
\mathrm{IT}_n & = & K_{Y_n+1}.\label{eq:generate_IT}
\end{eqnarray}

In \cite{Pareek:CNSNS2005}, it's said that DT2 has a number of
rows equal to the total number of session keys, which means that
the number of entries in DT2 is 16.

\item
The encryption process runs by reading each entry of DT2. For the
$n$-th entry, the chaotic map marked with number $N_n$ is chosen
to encrypt a group of $B_n$ plain-bytes. Each plain-byte $P_i$ is
masked by the chaotic state after $\mathrm{IT}_n$ iterations of
the chosen chaotic maps, according to the following rule:
\begin{equation}\label{eq:encryption}
C_i=\left(P_i+\lfloor X_{\mathrm{new}}\cdot
10^5\rfloor\right)\bmod 256.
\end{equation}
After each plain-byte is encrypted, IC of the chosen chaotic map
in DT1 is updated as $X_{\mathrm{new}}$. Once DT2 is exhausted,
substitute the latest value of IC in DT1 into Eq.~(\ref{eq:LCG0})
to reset $Y_0$, and then repeat Eqs. (\ref{eq:LCG}) to
(\ref{eq:generate_IT}) for 16 times to update all entries of DT2
for future encryption.

\item
The decryption procedure is similar to the above encryption
procedure, by replacing Eq.~(\ref{eq:encryption}) with the
following one:
\begin{equation}
P_i=\left(C_i-\lfloor X_{new}\cdot 10^5\rfloor\right)\bmod 256.
\end{equation}
\end{enumerate}

Wei et al. in \cite{JunWei:CNSNS2006} pointed out that the above
cipher works like a stream cipher, so a key-stream
$\left\{(C_i-P_i)\bmod 256\right\}$ can be constructed in
known-plaintext attacks and then be used as an equivalent of the
secret key $K$ to decrypt other ciphertexts. To overcome this
security problem, Wei et al. suggested a remedy to modify
Eq.~(\ref{eq:generate_N}), as follows:
\begin{equation}
N_n=(Y_n\bmod 4)\oplus\left(\bigoplus_{i=0}^{n-1}P_i\bmod
4\right),\label{eq:Wei_Nn}
\end{equation}
where $P_i$ is the $i$-th plain-byte, $\oplus$ denotes the bitwise
XOR operation, and $P_0=0$.

\section{Cryptanalysis}
\label{sec:cryptanalysis}

In addition to the defect of the original cipher of Pareek et al.
\cite{Pareek:CNSNS2005} pointed out in \cite{JunWei:CNSNS2006}, we
found some other security problems that exist in both the original
cipher and the improved version proposed in
\cite{JunWei:CNSNS2006}.

\subsection{Weak Keys}

Observing Eq.~(\ref{eq:generate_IC}), one can see the number of all
possible values of IC is only $256=2^8$, namely, $\frac{0}{256}\sim
\frac{255}{256}$. Since $x=0$ is a fixed point common to all the
four chaotic maps, $\mathrm{IC}=\frac{0}{256}$ will cause all
chaotic states to be zero, which means that $C_i=P_i$, $\forall i$.
In this case, the chaotic cipher does not work at all and the
corresponding key is an extremely weak key. To make
$\mathrm{IC}=\frac{0}{256}$, one has $\sum_{i=1}^{16}K_i\equiv
0\pmod{256}$. Then, one can calculate the number of such weak keys
to be $2^{16\times 8}/256=2^{15\times 8}=2^{120}$.
Figure~\ref{figure:WeakKey} shows the encryption result when a weak
key $K=61624D51595F888A434487885C5E483D$ (represented in hexadecimal
format) is used to encrypt a sinusoidal waveform.

Additionally, to ensure a higher level of security, the value of
IT should not be too small, which means that each sub-key should
not be too short. This will further reduce the key space.

\newlength\figwidth
\setlength\figwidth{0.9\columnwidth}

\begin{figure}[!htb]
\centering
\includegraphics[width=\figwidth]{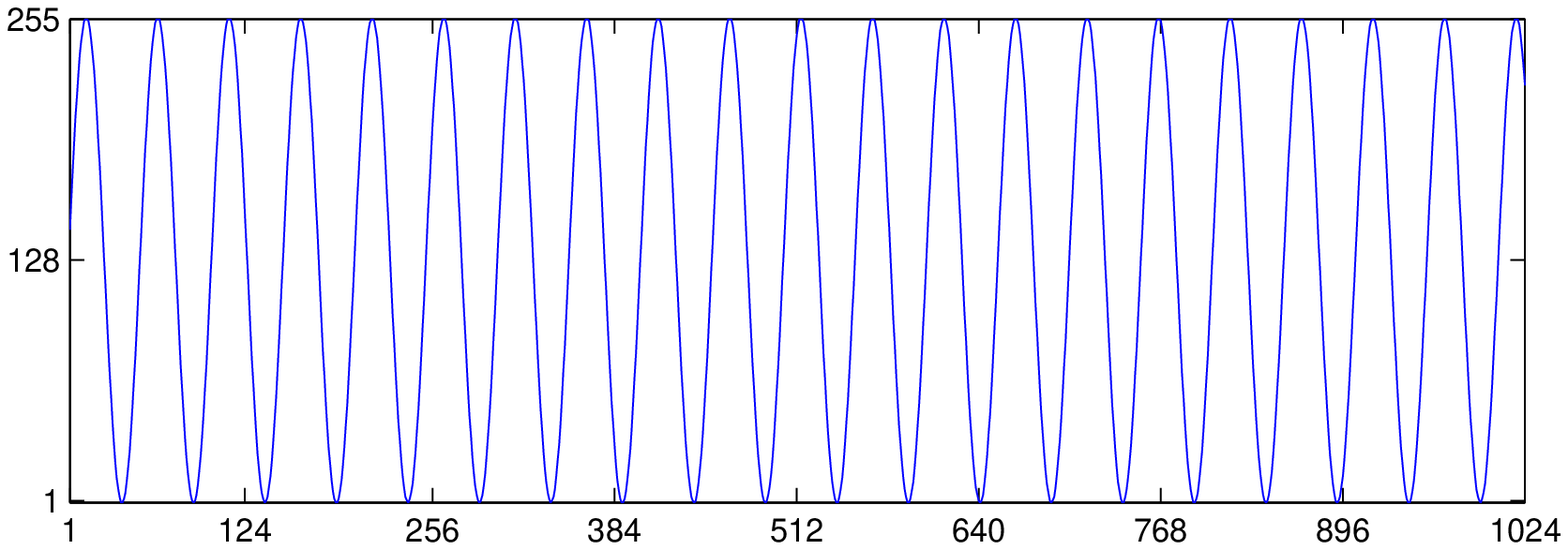}\\
a) the plaintext\\
\includegraphics[width=\figwidth]{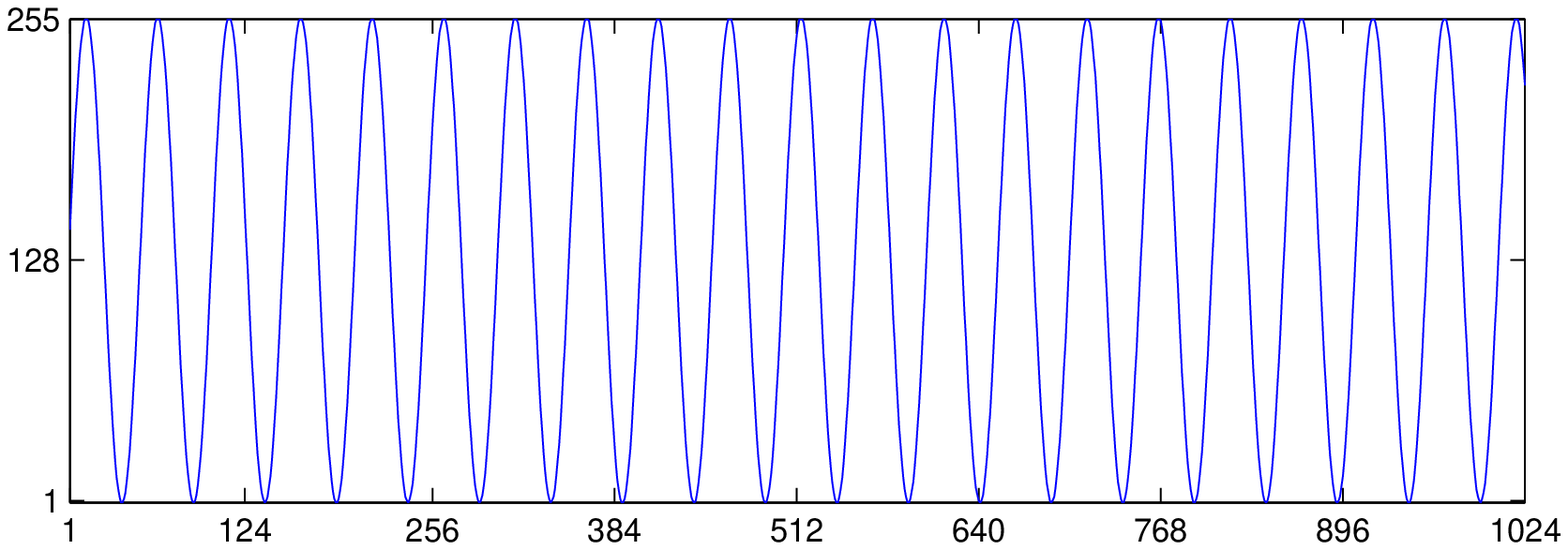}\\
b) the ciphertext\\
\caption{The encryption result of a sinusoidal waveform with one
weak key, ``61624D51595F888A434487885C5E483D".}
\label{figure:WeakKey}
\end{figure}

Finally, it is worth mentioning that the same kind of weak keys also
exists in the chaotic cipher proposed by Pareek et al. in
\cite{Pareek:PLA2003}, due to the similarity between the two
ciphers. This weakness had not been pointed out in Alvarez et al.'s
cryptanalysis paper \cite{Alvarez:PLA2003}.

\subsection{Weak Randomness of DT2}

The second dynamic table DT2 is generated in a pseudorandom way
using a LCG and controlled by the secret key. Such generators are
easy to implement and pass many statistical tests, thus leading to
believe that they are good candidates for generating strong
pseudorandom sequences for cryptographical applications. However,
these sequences are predictable: given a piece of the sequence, it
is possible to reconstruct all the rest even if the parameters are
unknown \cite{Boyar:LCG:JACM89}. Therefore, the use of linear
congruential generators in cryptography is totally discouraged.
Furthermore, the choice of parameters for the LCG in
\cite{Pareek:CNSNS2005} is most unfortunate. Using a prime number as
the modulus of the LCG would have yielded better results, but by
using $16$ as modulus, the randomness of its sequences is null. In
fact, the sequence is a unique cycle where the start value is the
seed of the LCG. The known-plaintext attack discussed in the next
subsection benefits from the lack of randomness of DT2, which
reduces the attacking complexity.

In the following, we prove some mathematical results on the LCG
sequence $\{Y_n\}$ and the map-number sequence $\{N_n\}$. It can
be seen that the two sequences are far from having ``good"
randomness.

\begin{lemma}\label{lemma}
Given a sequence $\{Y_n\}$, where $Y_n=(5Y_{n-1}+1)\bmod 16$ for
$n\geq 2$. We have $Y_n=\left(5^nY_0+\sum_{i=n-1}^05^i\right)\bmod
16$.
\end{lemma}
\begin{proof}
We prove this lemma via mathematical induction.

When $n=1$, $Y_1=(5Y_0+1)\bmod 16$, so the lemma is true.

Assuming $Y_n=\left(5^nY_0+\sum_{i=n-1}^05^i\right)\bmod 16$ holds
for $1\leq n\leq k$, we prove the lemma for the case of $n=k+1\geq
2$. From $Y_n=(5Y_{n-1}+1)\bmod 16$, we have
\begin{eqnarray*}
Y_{k+1} & = & (5Y_k+1)\bmod 16\\
& = & \left(5\left(\left(5^kY_0+\sum_{i=k-1}^05^i\right)\bmod
16\right)+1\right)\bmod 16\\
& = & \left(5\left(5^kY_0+\sum_{i=k-1}^05^i\right)+1\right)\bmod
16\\
& = & \left(5^{k+1}Y_0+\sum_{i=k}^05^i\right)\bmod 16.
\end{eqnarray*}
Thus, the lemma is proved.
\end{proof}

\begin{theorem}\label{theorem:Yn}
Given a sequence $\{Y_n\}_{n\geq 1}$, where $Y_n=(5Y_{n-1}+1)\bmod
16$ for $n\geq 2$. We have $Y_n=\left(2n^2+(4Y_0-1)n+Y_0\right)\bmod
16$.
\end{theorem}
\begin{proof}
From Lemma \ref{lemma}, we have
\begin{eqnarray*}
Y_n & = & \left(5^nY_0+\sum_{i=n-1}^05^i\right)\bmod 16=\left(5^nY_0+\frac{5^n-1}{5-1}\right)\bmod 16\\
& = & \left((1+4)^nY_0+\frac{(1+4)^n-1}{4}\right)\bmod 16\\
& = &
\left(\sum_{i=0}^n\binom{n}{i}4^iY_0+\frac{\sum_{i=0}^n\binom{n}{i}4^i-1}{4}\right)\bmod
16\\
& = & \left((1+4n)Y_0+\left(n+\binom{n}{2}4\right)\right)\bmod 16\\
& = & \left(2n^2+(4Y_0-1)n+Y_0\right)\bmod 16.
\end{eqnarray*}
This completes the proof of the theorem.
\end{proof}
\begin{corollary}\label{corollary:Yn16}
Given a sequence $\{Y_n\}_{n\geq 1}$, where $Y_n=(5Y_{n-1}+1)\bmod
16$ for $n\geq 2$. It has a period of $16$.
\end{corollary}
\begin{proof}
Assume the period of the sequence $\{Y_n\}$ is $T$. From Theorem
\ref{theorem:Yn}, we can get $Y_{n+16}-Y_n\equiv 0\pmod{16}$. This
means that $T|16$, i.e., $T\in\{1,2,4,8,16\}$. Again, from Theorem
\ref{theorem:Yn}, we have
\begin{eqnarray*}
Y_{n+8}-Y_n & \equiv &
\left(2(n+8)^2+(4Y_0-1)(n+8)+Y_0\right)\\
& & {}-\left(2n^2+(4Y_0-1)n+Y_0\right)\pmod{16}\\
& \equiv & 8\pmod{16}.
\end{eqnarray*}
Since $Y_n,Y_{n+8}\in\{0,\cdots,15\}$, the above result means
$Y_{n+8}\neq Y_8$. That is, $T>8\Rightarrow T=16$, which proves
the corollary.
\end{proof}
\begin{remark}
From Theorem \ref{theorem:Yn}, it is obvious that there are only 16
distinct sequences of $\{Y_n\}_{n\geq 1}$, shown as follows
($Y_1\sim Y_{16}$):\\\upshape
\[
\begin{array}{*{17}{c}}
1 & 6 & 15 & 12 & 13 & 2 & 11 & 8 & 9 & 14 & 7 & 4 & 5 & 10 & 3 & 0 & \cdots\\
6 & 15 & 12 & 13 & 2 & 11 & 8 & 9 & 14 & 7 & 4 & 5 & 10 & 3 & 0 & 1 & \cdots\\
15 & 12 & 13 & 2 & 11 & 8 & 9 & 14 & 7 & 4 & 5 & 10 & 3 & 0 & 1 & 6 & \cdots\\
12 & 13 & 2 & 11 & 8 & 9 & 14 & 7 & 4 & 5 & 10 & 3 & 0 & 1 & 6 & 15 & \cdots\\
13 & 2 & 11 & 8 & 9 & 14 & 7 & 4 & 5 & 10 & 3 & 0 & 1 & 6 & 15 & 12 & \cdots\\
2 & 11 & 8 & 9 & 14 & 7 & 4 & 5 & 10 & 3 & 0 & 1 & 6 & 15 & 12 & 13 & \cdots\\
11 & 8 & 9 & 14 & 7 & 4 & 5 & 10 & 3 & 0 & 1 & 6 & 15 & 12 & 13 & 2 & \cdots\\
8 & 9 & 14 & 7 & 4 & 5 & 10 & 3 & 0 & 1 & 6 & 15 & 12 & 13 & 2 & 11 & \cdots\\
9 & 14 & 7 & 4 & 5 & 10 & 3 & 0 & 1 & 6 & 15 & 12 & 13 & 2 & 11 & 8 & \cdots\\
14 & 7 & 4 & 5 & 10 & 3 & 0 & 1 & 6 & 15 & 12 & 13 & 2 & 11 & 8 & 9 & \cdots\\
7 & 4 & 5 & 10 & 3 & 0 & 1 & 6 & 15 & 12 & 13 & 2 & 11 & 8 & 9 & 14 & \cdots\\
4 & 5 & 10 & 3 & 0 & 1 & 6 & 15 & 12 & 13 & 2 & 11 & 8 & 9 & 14 & 7 & \cdots\\
5 & 10 & 3 & 0 & 1 & 6 & 15 & 12 & 13 & 2 & 11 & 8 & 9 & 14 & 7 & 4 & \cdots\\
10 & 3 & 0 & 1 & 6 & 15 & 12 & 13 & 2 & 11 & 8 & 9 & 14 & 7 & 4 & 5 & \cdots\\
3 & 0 & 1 & 6 & 15 & 12 & 13 & 2 & 11 & 8 & 9 & 14 & 7 & 4 & 5 & 10 & \cdots\\
0 & 1 & 6 & 15 & 12 & 13 & 2 & 11 & 8 & 9 & 14 & 7 & 4 & 5 & 10 & 3 & \cdots\\
\end{array}
\]
\it It can be seen that the 16 sequences actually represent the same
sequence with different starting points. This is a common feature of
discrete maps defined over a finite field and with a maximal period
\cite{Robert:Discrete-Iterations86}.
\end{remark}
\begin{corollary}\label{corollary:Yn4i}
Given a sequence $\{Y_n\}_{n\geq 1}$, where $Y_n=(5Y_{n-1}+1)\bmod
16$ for $n\geq 2$. Then, for any $n\geq 0$,
$\{Y_n,Y_{n+4},Y_{n+8},Y_{n+12}\}$ must be one of the following four
sets: $\{0,12,8,4\}$, $\{1,13,9,5\}$, $\{2,14,10,6\}$ and
$\{3,15,11,7\}$.
\end{corollary}
\begin{proof}
From Theorem \ref{theorem:Yn},
$Y_{n+4}-Y_n\equiv\left(2(n+4)^2+(4Y_0-1)(n+4)+Y_0\right)-\left(2n^2+(4Y_0-1)n+Y_0\right)\equiv
(4Y_0-1)4\equiv-4\pmod{16}$. Since $Y_n\in\{0,1,\cdots,15\}$, the
corollary is immediately proved. (The corollary can also be proved
by exhaustively examining all 16 distinct sequences of $\{Y_n\}$.)
\end{proof}
\begin{theorem}\label{theorem:Nn}
Given a sequence $\{Y_n\}_{n\geq 1}$, where $Y_n=(5Y_{n-1}+1)\bmod
16$ for $n\geq 2$. Then, assuming $N_n=Y_n\bmod 4$, we have
$N_n=(n+Y_0)\bmod 4$.
\end{theorem}
\begin{proof}
Substituting the result of Theorem \ref{theorem:Yn} into
$N_n=Y_n\bmod 4$, we have $N_n=Y_n\bmod
4=\left(2n^2+(4Y_0-1)n+Y_0\right)\bmod 4=(2n^2-n+Y_0)\bmod 4$. Note
that $(2n^2-n)-n\equiv 2n(n-1)\equiv 0\pmod 4$, so $2n^2-n\equiv
n\pmod 4$. This immediately leads to $N_n=(n+Y_0)\bmod 4$ and proves
the theorem.
\end{proof}
\begin{corollary}\label{corollary:Nn}
Given two sequences $\{Y_n\}_{n\geq 1}$ and $\{N_n\}_{n\geq 1}$,
where $Y_n=(5Y_{n-1}+1)\bmod 16$ for $n\geq 2$ and $N_n=Y_n\bmod 4$.
Then, the sequence $\{N_n\}_{n\geq 1}$ has a periodicity of 4, and
must be one of the following four sequences: $\{1,2,3,0,\cdots\}$,
$\{2,3,0,1,\cdots\}$, $\{3,0,1,2,\cdots\}$ and $\{0,1,2,3,\cdots\}$.
\end{corollary}
\begin{proof}
This corollary is a straightforward consequence of Theorem
\ref{theorem:Nn}.
\end{proof}

\subsection{Breaking the Secret Key by a Known-Plaintext Attack}

In \cite[Sec.~4]{JunWei:CNSNS2006}, Wei et al. pointed out that the
original cipher of Pareek et al. is vulnerable to known-plaintext
attacks. However, Wei et al.'s attack does not break the secret key
itself, but only reveals an equivalent of the secret key -- the key
stream $\{(C_i-P_i)\bmod 256=\lfloor X_{\mathrm{new},i}\cdot
10^5\rfloor\bmod 256\}$. The main disadvantage of this attack is
that it can only break a ciphertext as long as the keystream
recovered. In the real world, this means than long messages might
not be broken if a previous message just as long is not known.

In this section, we report a practical known-plaintext attack to
completely reveal the secret key, with only 120 consecutive known
plain-bytes in just one known plaintext, with rather small
computational complexity. This attack is very practical in real
world scenarios.

From Corollary~\ref{corollary:Nn}, one can see that for all
$n\in\{1, 2, 3, 4\}$, the plain-bytes in the $n$, $(n+4)$, $(n+8)$,
$(n+12)$-th groups are encrypted by the chaotic map numbered with
$N_n=N_{n+4}=N_{n+8}=N_{n+12}$. At the same time, from
Corollary~\ref{corollary:Yn16}, the 16 ITs in DT2 form a permutation
of the 16 sub-keys $K_1$, $\cdots$, $K_{16}$. The two facts mean
that we can try to \textbf{separately} break the sub-keys used for
each chaotic map. If such a divide-and-conquer (DAC) attack really
works, the total complexity of revealing all 16 sub-keys will be
dramatically reduced as compared with exhaustively searching them
throughout the whole key space.

It is found that a three-stage DAC attack shown below works well
following the above idea.
\begin{itemize}
\item
\textit{Stage 1 -- exhaustively guessing \textup{IC} in
Eq.~(\ref{eq:generate_IC}) and 4 sub-keys (i.e., \textup{IT}s)
used by one chaotic map numbered with $N_n$.}

For each guessed value of IC, the chaotic map is chosen to ensure
that $\{B_n,B_{n+4},B_{n+8},B_{n+12}\}$ does not contain
zero\footnote{Corollary~\ref{corollary:Nn} ensures that there are
always three chaotic maps of this kind. We can randomly choose one
from the three.}. To eliminate incorrectly guessed values of IC,
the repeated use of $\mathrm{IT}_n$ in each group is employed --
all $B_n$ chaotic states in the $n$-th group should correspond to
the same value of $\lfloor X_{\mathrm{new}}\times 10^5\rfloor\bmod
256=(C_i-P_i)\bmod 256$.

The output of this stage will be some candidate values of IC, each
of which corresponds to 4 revealed sub-keys. Without loss of
generality, assume that the chaotic map has a uniform invariant
distribution. Then, we can calculate the probability of getting a
wrong candidate value
\[
P_e=\dfrac{256^4}{256^{B_n+B_{n+4}+B_{n+8}+B_{n+12}}}.
\]
It follows from Corollary~\ref{corollary:Nn} that
\[
P_e\leq\dfrac{256^4}{256^{1+13+9+5}}=256^{-24}=2^{-192}.
\]
To further minimize the value of $P_e$, for each guessed value of
IC, one can chose the map corresponding to
$\{B_n,B_{n+4},B_{n+8},B_{n+12}\}=\{3,15,11,7\}$. In this way, $P_e$
will be minimized to be $256^4/256^{3+15+11+7}=256^{-32}=2^{-256}$.
Thus, it is an extremely rare event to get more than one candidate
value of IC in practice\footnote{Even when such a rare event
happens, one can verify all the candidate values by choosing another
chaotic map. This will further eliminate wrong candidate values and
eventually leave only the correct one.}.

\item
\textit{Stage 2 -- exhaustively searching other 11 sub-keys (i.e.,
\textup{IT}s) used by other three chaotic maps.}

Once the value of IC is determined, we can use a similar method in
Stage 1 to determine the sub-keys used by other three chaotic
maps. Note that the sub-key corresponding to $B_n=0$ cannot be
found, since no any plain-byte is encrypted with this sub-key. So,
only 11 sub-keys can be revealed in this stage and the last one is
left for the next stage.

\item
\textit{Stage 3 -- revealing the last unknown sub-key via
Eq.~(\ref{eq:generate_IC}).}

In the above two stages, one can successfully get the value of IC
and break 15 sub-keys. The last sub-key can be determined via
Eq.~(\ref{eq:generate_IC}). Assuming the undetermined sub-key is
$K_j$, we have
\begin{equation}
K_j=\left(256\times\mathrm{IC}-\sum_{1\leq i\leq 16 \atop i\neq
j}K_i\right)\bmod 256.
\end{equation}
\end{itemize}

Now, let us estimate the computational complexity of this attack.
First, the computational complexity of Stage 3 is very small, so
we can consider only the first two stages. By enumerating the
number of guessed values of IC and the number of all chaotic
iterations, we can deduce that the computational complexity of
Stage 1 is not greater than $O(255\times
256\times(3+15+11+7))\approx O(2^{21})$ and Stage 2 is not greater
than $O(256\times(3+15+11+7+2+14+10+6+0+12+8+4))\approx
O(2^{14.5})$. As a whole, the total complexity of the DAC attack
is mainly determined by Stage 1, which is not greater than
$O(2^{21})$. Attacks with such small complexity can be easily
carried out on a PC.

Besides the very small computational complexity, the required
number of known plain-bytes in an attack is also very small --
only $\sum_{i=1}^{16}B_i=\sum_{i=1}^{16}i=120$ plain-bytes in one
known plaintext are enough.

\begin{figure}[!htb]
\centering
\includegraphics[width=\figwidth]{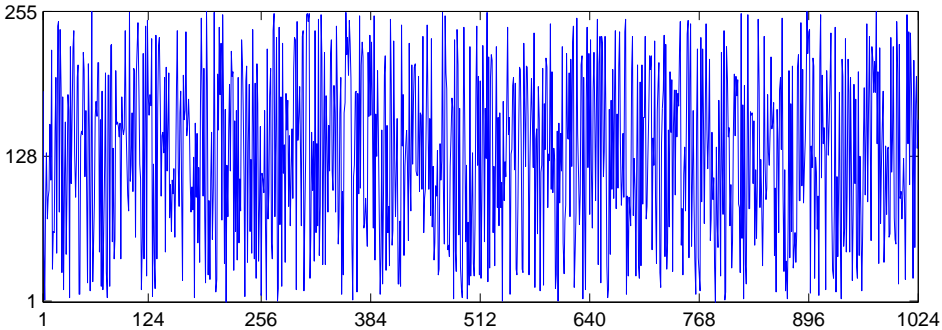}
\caption{The ciphertext of the sinusoidal waveform shown in
Fig.~\ref{figure:WeakKey}a), with
$K=BCDA178E512131422E859F086E2E884F$.} \label{figure:Ciphertext}
\end{figure}

The above analysis shows that the proposed DAC attack is very
efficient. To further validate the feasibility of the attack, a real
attack was carried out with one known plaintext as shown in
Fig.~\ref{figure:WeakKey}a) and the corresponding ciphertext shown
in Fig.~\ref{figure:Ciphertext}. The breaking results obtained in
all the three stages are given in Table~\ref{table:example}. With
the broken sub-keys, one can immediately get the whole secret key,
$K=K_1\cdots K_{16}=BCDA178E512131422E859F086E2E884F$ (represented
in hexadecimal format).

\begin{table}[!htbp]
\center \caption{The stage-by-stage breaking results of a real
example of the proposed known-plaintext
attack.}\label{table:example}
\begin{tabular}{|c|*{3}{c|}}
\hline & Stage 1 & Stage 2 & Stage 3\\
\hline\hline IC & $\frac{237}{256}$ & &\\
\hline $K_{14}=\mathrm{IT}_1$ & 146 & &\\
\hline $K_3=\mathrm{IT}_2$ & & 23 &\\
\hline $K_{12}=\mathrm{IT}_3$ & & 8 &\\
\hline $K_9=\mathrm{IT}_4$ & & 46 &\\
\hline $K_{10}=\mathrm{IT}_5$ & 133 & &\\
\hline $K_{15}=\mathrm{IT}_6$ & & 136 &\\
\hline $K_8=\mathrm{IT}_7$ & & 66 &\\
\hline $K_5=\mathrm{IT}_8$ & & 81 &\\
\hline $K_6=\mathrm{IT}_9$ & 33 & &\\
\hline $K_{11}=\mathrm{IT}_{10}$ & & 159 &\\
\hline $K_4=\mathrm{IT}_{11}$ & & 142 &\\
\hline $K_1=\mathrm{IT}_{12}$ & & & 188\\
\hline $K_2=\mathrm{IT}_{13}$ & 218 & &\\
\hline $K_7=\mathrm{IT}_{14}$ & & 49 &\\
\hline $K_{16}=\mathrm{IT}_{15}$ & & 79 &\\
\hline $K_{13}=\mathrm{IT}_{16}$ & & 110 &\\
\hline
\end{tabular}
\end{table}

\subsection{Security Problem of Wei et al.'s Version}

The improved version of the original cipher, proposed by Wei et
al. in \cite{JunWei:CNSNS2006}, employs plaintext feedback to
enhance the security against the simple keystream-based
known-plaintext attack. However, even this cipher cannot resist
the DAC attack proposed-above in this paper, because this attack
does not depend on the relation between the keystream and the
plaintext. Of course, in the cipher of Wei et al., because the
periodicity of $\{N_n\}_{n\geq 1}$ is destroyed by the plaintext
feedback, the performance of the DAC attack may be complicated
slightly. The main influence includes the following two aspects.

First, in Stage 1, the plaintext feedback influences the manner of
choosing the target chaotic map, since now the $n$-th chaotic map
generally does not correspond to
$\{B_n,B_{n+4},B_{n+8},B_{n+12}\}$, but to a set
$\{B_{n_1},B_{n_2},\cdots,B_{n_i}\}$ whose size depends on the
plaintext. To minimize the value of $P_e$, we should choose the
target chaotic map as the one with the maximal value of
$\sum_{j=1}^{i}B_{n_i}$. Since
$\sum_{j=1}^{16}B_j=\sum_{j=1}^{16}(j-1)=120$, we can deduce
$\sum_{j=1}^{i}B_{n_i}\geq 120/4=30$. This means that
$P_e\leq\frac{256^4}{256^{30}}=256^{-26}=2^{-208}$. So, it is
still an extremely rare event to get more than one candidate value
after Stage 1 is completed.

Second, in Stage 2, for one or two chaotic maps, the value of
$\sum_{j=1}^{i}B_{n_i}$ may not be large enough to uniquely
determine the values of some sub-keys. In this case, only 120
plain-bytes will not be enough to recover all sub-keys.
Nevertheless, the probability of this event is not too
large\footnote{It is not easy to theoretically deduce this
probability. Assuming all chaotic maps satisfy $P_e\leq 10^{-4}$,
we found the probability is not greater than 0.06 with 300,000
random experiments in Matlab.}, so these undetermined sub-keys
will be gradually broken with the accumulation of more known
plain-bytes.

Finally, the following two points on the security of Wei et al.'s
improved cipher are worth mentioning: 1) in the chosen-plaintext
counterpart of the DAC attack, the plaintext feedback mechanism
can be completely circumvented by choosing all plain-bytes to be
zero; 2) the plaintext feedback cannot rule out the existence of
weak keys and the weak randomness of $\{B_n\}_{n\geq 1}$. To sum
up, Wei et al.'s remedy is not essentially improving the security
of the original cipher of Pareek et al.

\section{Conclusions}

In this paper, the security of a recently-proposed cipher based on
multiple one-dimensional chaotic maps \cite{Pareek:CNSNS2005} has
been re-examined, showing that a previous cryptanalysis
\cite{JunWei:CNSNS2006} did not reveal many major security
problems. As a result, a number of weak keys and weak
pseudorandomness of some intermediate data were discovered and
distinguished, and an efficient known-plaintext attack can be
recommended to completely reveal the whole secret key. The
proposed attack has a very small computational complexity, which
works with only 120 plain-bytes in one known plaintext. In
addition, it is found that an improved version of the original
cipher, proposed in \cite{JunWei:CNSNS2006}, also suffers from the
same security problems. The cryptanalysis given in this paper thus
discourages the use of the chaotic cipher proposed in
\cite{Pareek:CNSNS2005, JunWei:CNSNS2006}, especially when
known-plaintext attacks are possible.

\begin{ack}
This research was partially supported by The Hong Kong Polytechnic
University's Postdoctoral Fellowships Program under grant no. G-YX63
and by Ministerio de Ciencia y Tecnologia of Spain, research grant
SEG2004-02418. The work of K.-T. Lo was supported by the Research
Grants Council of the Hong Kong SAR Government under Project Number
523206 (PolyU 5232/06E).
\end{ack}

\bibliographystyle{elsart-num}
\bibliography{CNSNS}

\end{document}